\documentclass[onecolumn,prb]{revtex4}
\usepackage{amsfonts, amsmath, latexsym}
\usepackage[dvips]{graphicx}

\newcommand{\bra}[1]{\langle #1|}
\newcommand{\ket}[1]{|#1\rangle}

\begin{document}

\title{Isomorphic classical molecular dynamics model for an excess electron in a supercritical fluid}
\author{Thomas F. Miller, III} \email{tfm@caltech.edu}
 \affiliation{Department of Chemistry, University of California, Berkeley, CA 94720}
 \altaffiliation[Current address: ]{Arthur Amos Noyes Laboratory of Chemical Physics, California Institute of Technology, Pasadena, California 91125 }
 
\vspace*{2cm}
\maketitle

\noindent
{\bf ABSTRACT}

\noindent

Ring polymer molecular dynamics (RPMD) is used to directly simulate the dynamics of an excess electron in a supercritical fluid over a broad range of densities. 
The accuracy of the RPMD model is tested against numerically exact path integral statistics through the use of analytical continuation techniques.  At low fluid densities, the RPMD model substantially underestimates the contribution of delocalized states to the dynamics of the excess electron.  However, with increasing solvent density, the RPMD model improves, nearly satisfying analytical continuation constraints at densities approaching those of typical liquids.  In the high density regime, quantum dispersion substantially decreases the self-diffusion of the solvated electron. 
  In this regime where  the dynamics of the electron is strongly coupled to the dynamics of the atoms in the fluid,  trajectories that can reveal diffusive motion of the electron are long in comparison to $\beta\hbar$.  
\newpage

\section{Introduction}


Ring polymer molecular dynamics (RPMD) allows for the direct simulation of quantum mechanical systems. 
It is a classical model that both preserves the exact quantum Boltzmann distribution and exhibits time-reversal symmetry.\cite{Cra04,Bra06}  These features ensure that  RPMD trajectories are stable and self-consistent on long timescales, enabling the study of coupled dynamical timescales in complex systems.  In exhibiting these features, RPMD and centroid molecular dynamics\cite{Cao94,Jan99} are unique among quantum dynamical methods.  
Mixed quantum-classical methods based on mean-field\cite{McL64,Mey80} and trajectory surface hopping\cite{Tul90,Tul98} dynamics do not preserve the correct Boltzmann populations.\cite{Nie01,Par05,Sch08}
Similarly, methods based on the classical Wigner model\cite{Wan98,Sun98,Her98,Pol98,Sha99,Shi03,Pou03,Liu07}  do not yield trajectories that  preserve the quantum Boltzmann distribution and are thus not employed as models for direct dynamical simulation.



Here, we test the RPMD model for the direct simulation of an excess electron in a supercritical fluid.  This is the first application of RPMD to electronic degrees of freedom, a highly quantum mechanical - and presumably challenging - regime for the model.  
To evaluate the accuracy of RPMD in this context, we compare the model dynamics against numerically exact path integral statistics with the aid of analytical continuation techniques.  We further explore the long-timescale features of the dynamics using direct simulation with RPMD.

\section{Methods}

 
 An excess electron is simulated in an otherwise classical fluid.  The system is described using potentials and parameters
 that were previously adopted by Berne, Coker, and coworkers to model 
 an excess electron in helium at $300$ K.\cite{Hir54,Kes65,Cok87,Cok88,Spa91,Spa92,Gal94,Gal96}
 The potential energy function, 
 $U({\bf q},{\bf Q}_1,\ldots,{\bf Q}_N)$, is a sum of pairwise helium-helium and electron-helium interactions;\cite{Cok87}
${\bf q}$ and ${\bf Q}_j$ specify the positions of the electron and the helium atoms, and $N$ is the number of atoms in the fluid.  
Helium-helium interactions are described using the Lennard-Jones potential with $\sigma_{\textrm{He}}=2.556$ \AA\ and $\epsilon_{\textrm{He}} = 10.22$ K; electron-helium interactions are described using the pairwise pseudopotential
 \begin{equation}
 U_{\textrm{e-He}}(r)=\frac{A}{r^4}\left[\frac{B}{\left(C+r^6\right)}-1\right],
\label{Vehe}
 \end{equation}
 where $r$ is the electron-helium distance and, in atomic units, $A=0.655$, $B=89099$, and $C=12608$.

The ring polymer molecular dynamics (RPMD) equations of motion for this system are\cite{Cra04}
\begin{eqnarray}
\dot{{\bf v}}^{(\alpha)} &=& \omega_n^{2}({\bf q}^{(\alpha+1)}+{\bf q}^{(\alpha-1)}-2{\bf q}^{(\alpha)})-\frac{1}{m}\nabla_{{\bf q}^{(\alpha)}} U({\bf q}^{(\alpha)},{\bf Q}_1,\ldots,{\bf Q}_N)\nonumber\\
\dot{{\bf V}}_j &=& -\frac{1}{nM}\sum_{\alpha=1}^n \nabla_{{\bf Q}_j} U({\bf q}^{(\alpha)},{\bf Q}_1,\ldots,{\bf Q}_N),
\label{mixedEOM}
\end{eqnarray}
where the integer $\alpha$ indexes the ring polymer beads for the electron, such that ${\bf q}^{(0)}={\bf q}^{(n)}$.  Also, ${\bf v}^{(\alpha)}$ and ${\bf V}_j $ specify the respective velocities for the ring polymer beads and the classical helium atoms, $m$ and $M$ are the respective electron and helium masses, and $\omega_n=n(\beta\hbar)^{-1}$, where $\beta$ is the reciprocal temperature. 
%
These dynamics preserve the path integral discretization for the Boltzmann distribution,\cite{Fey65,Cha81}
\begin{equation}
P(\{{\bf q}^{(\alpha)}\},{\bf Q}_1,\ldots,{\bf Q}_N)
\propto
\textrm{exp} \left\{-
\frac{\beta}{n}\sum_{\alpha=1}^n\left( \frac12 m\omega_n^2({\bf q}^{(\alpha)}-{\bf q}^{(\alpha-1)})^2+U({\bf q}^{(\alpha)},{\bf Q}_1,\ldots,{\bf Q}_N)\right)\right\},
\end{equation}
such that in the limit of large $n$, sampling the RPMD trajectories yields exact equilibrium averages.\cite{Par84,Rae84} 
For example, the imaginary time mean-squared displacement\cite{Cha90} 
\begin{equation}
R^2(\tau)=\langle |{\bf q}(\tau)-{\bf q}(0)|^2\rangle,
\label{eq:ctim}
\end{equation}
is obtained from the RPMD using
\begin{equation}
R^2(\tau_\alpha)=
\left\langle({\bf q}^{(\alpha)}-{\bf q}^{(0)})^2\right\rangle_{\textrm{RP}}=
\frac{1}{T_{\textrm{obs}}}\sum_{s=1}^{T_{\textrm{obs}}} \left({\bf q}^{(\alpha)}(s) - {\bf q}^{(0)}(s)\right)^2,
\label{pimd}
\end{equation}
where $\tau_\alpha=j\beta\hbar/n$, and ${\bf q}^{(\alpha)}(t)$ indicates the position of bead $\alpha$ after evolution for time $t$ using  Eq. (\ref{mixedEOM}).  $T_{\textrm{obs}}$ is the total length of the RPMD trajectory, which must be thermostatted to be fully ergodic.
For the calculation of static quantities, RPMD is equivalent to the path integral molecular dynamics (PIMD) method.\cite{Par84,Rae84} 


Beyond calculating equilibrium properties, RPMD utilizes the time-displaced statistics to estimate real-time behavior.
It has been employed previously as a method to calculate Kubo-transformed correlations functions,\cite{Cra04,Mil05a,Mil05b,Cra05a,Cra05b}
 such as the velocity autocorrelation function to which the mobility of the excess electron and its absorption spectrum can be related,
\begin{equation}
\tilde{c}_{{\bf v}\cdot {\bf v}}(t)=\frac{1}{m^2\beta}\int^\beta_0\ d\lambda\ 
\langle {\bf p}(-i\lambda\hbar)\cdot {\bf p}(t)\rangle.
\end{equation}
Here, ${\bf p}=-i\hbar\nabla_{\bf q}$ is the momentum operator for the electron.
The RPMD approximation to this correlation function is\cite{Mil05a,Mil05b}
\begin{equation}
\tilde{c}_{{\bf v}\cdot {\bf v}}(t) \approx
\left\langle
\bar{{\bf v}}(0)\cdot\bar{{\bf v}}(t)
\right\rangle_{\textrm{RP}}
 = \frac{1}{T_{\textrm{obs}}}\sum_{s=1}^{T_{\textrm{obs}}} {\bar{\bf v}}(s)\cdot\bar{{\bf v}}(s+t),
\label{ctvv}
\end{equation}
where
$\bar{{\bf v}}(t)$ indicates the bead-averaged momentum for the electron ring polymer that is also evolved according to Eq. (\ref{mixedEOM}).  

In the current work, we take a somewhat expanded view of the RPMD model, using it to directly simulate the time-evolution of a quantum system.  For example, rather than calculating the self-diffusion coefficient of the excess electron from the integral of $\tilde{c}_{{\bf v}\cdot {\bf v}}(t)$, we instead simulate the real-time displacement of the electron using
\begin{equation}
R^{2}(t) \approx 
\left\langle(\bar{{\bf q}}(t)-\bar{{\bf q}}(0))^2
\right\rangle_{\textrm{RP}},
\label{ctr2}
\end{equation}
where $\bar{{\bf q}}(t)$ is the bead-averaged position for the time-evolved electron ring polymer.
The RPMD model obeys the classical molecular dynamics relationship\cite{All87}
\begin{equation}
\frac12\frac{d}{dt}R^{2}(t) = \int_0^t dt' \ \tilde{c}_{{\bf v}\cdot {\bf v}}(t'),
\label{r2vsctvv}
\end{equation}
so that both approaches yield the same estimate for the self-diffusion coefficient.\cite{centroid_einstein}  As another example, we note that the RPMD estimate for the thermal rate constant need not be obtained from the calculation of a time correlation function; an equivalent estimate would be obtained by simply running a long RPMD trajectory and counting the frequency with which the reaction of interest occurs.\cite{Col08}

At the heart of these and other robust features of RPMD is the status of the model as a genuine classical dynamics, albeit one that preserves the quantum Boltzmann distribution.  RPMD ensures time reversibility, and it exhibits time-translational invariance at equilibrium. 
 It is a model for simulating the real-time evolution of a system, in addition to a means of calculating time correlation functions.  RPMD can thus be employed outside of the time correlation function formalism and beyond the linear response regime.  In the same way that standard molecular dynamics is used to simulate classical reaction mechanisms and to simulate classical processes far from equilibrium, we propose that RPMD be used to simulate dynamics in systems for which quantum statistical effects play an important role.  The centroid molecular dynamics model shares these features and prospects, although it has not, to our knowledge, fully utilized this philosophy.

\subsection{Relating static and real-time correlation functions via analytical continuation}

To evaluate the accuracy of RPMD for the dynamics of the solvated electron, we will relate the results of the model dynamics to numerically exact path integral statistics through the use of analytical continuation.\cite{Bay61,Gub91}  This approach is described in the next two sections.

The dipole autocorrelation function for the solvated electron is
\begin{equation}
C(t)=\langle {\bf \mu}(t)\cdot {\bf \mu}(0)\rangle=
\frac{1}{Z}\textrm{Tr}(e^{-\beta \hat{H}} e^{it\hat{H}/\hbar}\mu e^{-it\hat{H}/\hbar}\cdot\mu).
\end{equation}
Here, $Z=\textrm{Tr}(e^{-\beta \hat{H}})$ is the canonical partition function, $\mu=-|e|{\bf q}$ is the dipole operator, $e$ is the charge of an electron, and $\beta=1/(k_BT)$ is the reciprocal temperature. 
The dipole spectral density function, $I(\omega)$, is defined such that
\begin{equation}
I(\omega)= \int_{-\infty}^{\infty}dt\ e^{i\omega t}C(t)\qquad\textrm{and}\qquad
C(t)= \frac{1}{2\pi}\int_{-\infty}^{\infty}d\omega\ e^{-i\omega t}I(\omega).
\label{ftransform}
\end{equation} 

The correlation function $C(t-i\tau)$ is analytic in the strip $0\le\tau\le\beta\hbar$, where $t$ and $\tau$ are real numbers.  We can thus introduce the imaginary-time correlation function $G(\tau)\equiv C(-i\tau)$ such that, by way of analytic continuation, $G(\tau)$ formally contains all of the dynamical information on the real-time axis.\cite{Bay61}  

It follows from Eq. (\ref{ftransform}) that $G(\tau)$ and $I(\omega)$ are related by a double-sided Laplace transform,
\begin{equation}
G(\tau)= \frac{1}{2\pi}\int_{-\infty}^{\infty}d\omega\ e^{-\omega \tau}I(\omega).
\end{equation}
By discretizing $G(\tau)$ in imaginary time, it is then straightforward to show that
\begin{eqnarray}
R^2(\tau) &=& \frac{2}{e^2}(G(0)-G(\tau))\nonumber\\
&=& \frac{1}{\pi e^2} \int_{-\infty}^{\infty} d\omega\ I(\omega)(1-e^{-\omega\tau}),
\end{eqnarray} 
where $R^2(\tau)$ is the mean squared displacement of the electron in imaginary time from Eq. (\ref{eq:ctim}).
Finally, using the detailed balance relation $I(-\omega)=e^{-\beta\hbar\omega}I(\omega)$, we obtain
\begin{equation}
R^2(\tau)=\frac{\hbar c}{4\pi^2e^2}\int_0^{\infty} d\omega\ K(\tau,\omega) \sigma(\omega),
\label{inveq}
\end{equation}
where $c$ is the speed of light.  Here, we have introduced the absorption cross section
\begin{equation}
\sigma(\omega) = \frac{4\pi}{\hbar c}\omega(1-e^{-\beta\hbar\omega})I(\omega),
\end{equation}
and
\begin{equation}
K(\omega,\tau)=\frac{\textrm{cosh}(\beta\hbar\omega/2)-\textrm{cosh}(\beta\hbar\omega/2-\tau\omega)}{\omega\ \textrm{sinh}(\beta\hbar\omega/2)}.
\label{Kdef}
\end{equation}
Eqs. (\ref{inveq})-(\ref{Kdef}) were previously reported by Gallicchio and Berne.\cite{Gal94,Gal96}

\subsection{Maximum Entropy Analytical Continuation}
\label{sec:meac}
The analytical continuation result in Eq. (\ref{inveq}) suggests an integral inversion problem that yields real-time dynamics ($\sigma(\omega)$) from purely statistical information ($R^2(\tau)$).  However, this inversion is known to be numerically unstable, such that small errors in $R^2(\tau)$ lead to large errors in $\sigma(\omega)$.  Nonetheless, the inversion can be performed indirectly with the aid of information theory methods such as the maximum entropy technique.\cite{Gub91,Ski84,Buc91,Jar96}
  Maximum entropy analytical continuation (MEAC) 
yields $\sigma(\omega)$ by refining a prior estimate for the spectrum, $\sigma^\circ(\omega)$, against a numerical calculation for $R^2(\tau)$.  In the current study, we shall obtain $\sigma^\circ(\omega)$ from the RPMD model, and we shall obtain $R^2(\tau)$ from numerically exact path integral statistics. The degree to which the maximum entropy inversion alters the RPMD prior provides insight to the accuracy of the RPMD model for the excess electron.
The idea of using MEAC to correct approximate RPMD correlation functions was recently put forward by Habershon \textit{et al}.,\cite{Hab07} and MEAC was previously applied to combine centroid molecular dynamics with imaginary time data.\cite{Kri99}  Our implementation of the maximum entropy inversion follows that of Habershon \textit{et al}.\cite{Hab07} and Gallicchio and Berne,\cite{Gal94,Gal96} which are in turn based on the
prescription provided by Bryan.\cite{Bry90}

In our numerical implementation, $R^2(\tau)$ is known on the discrete set of $n$ points $\{\tau_j\}$, and we wish to solve for $\sigma(\omega)$ on the discrete set of $\mathcal{N}$ points $\{\omega_k\}$.  Eq. (\ref{inveq}) is thus expressed as a matrix equation
\begin{equation}
\textbf{c}=\textbf{K}\textbf{s}
\end{equation}
where $\textbf{c}$ is a column vector with elements $\textrm{c}_j=R^2(\tau_j)$, $\textbf{K}$ is a rectangular matrix with elements $K_{jk}=K(\tau_j,\omega_k)$, $\textbf{s}$ is a column matrix with elements $s_k=\sigma(\omega_k)\Delta\omega_k$, and $\Delta\omega_k$ is a quadrature weight.
 The maximum entropy approach does not seek to directly invert this equation, but rather to maximize the scoring function
 \begin{equation}
 Q(\textbf{s}; \alpha)=\alpha S(\textbf{s})-\chi^2(\textbf{s})/2
 \label{Qopt}
 \end{equation}
 with respect to the elements of $\textbf{s}$.  In this equation, $\chi^2(\textbf{s})$ is a measure of the degree to which $\textbf{s}$ fits the imaginary time data, $S(\textbf{s})$ is a measure of the degree to which $\textbf{s}$ deviates from an initial guess for the solution, and $\alpha$ is a parameter that tunes the relative influence of these measures during maximization. 
 
  Specifically,
 \begin{equation}
 \chi^2(\textbf{s}) = (\textbf{c}-\textbf{Ks})^T\textbf{C}^{-1}(\textbf{c}-\textbf{Ks}),
 \end{equation}
 where $\textbf{C}$ is the covariance matrix
 \begin{equation}
 C_{ij}=\frac{1}{\mathcal{M}(\mathcal{M}+1)}\sum_{k=1}^\mathcal{M}
 \left(\langle c_i\rangle-c_i^{(k)}\right)
 \left(\langle c_j\rangle-c_j^{(k)}\right),
 \label{covar}
 \end{equation} 
 and $\mathcal{M}$ is the number of statistically independent measurements of the imaginary time correlation function $R^2(\tau_j)$ obtained during the equilibrium sampling of the quantum Boltzmann distribution.  
 
 Given a prior model for the spectrum $\sigma^\circ(\omega_k)$, 
 we can define the information entropy
 \begin{equation}
 S(\textbf{s})=\sum_{k=1}^{\mathcal{N}}\left[s_k-s^\circ_k-s_k\textrm{ln}\frac{s_k}{s^\circ_k}\right],
 \label{entropy}
 \end{equation}
 where $s^\circ_k=\sigma^\circ(\omega_k)\Delta\omega_k$.  
 To obtain the RPMD prior spectrum $\sigma^\circ(\omega)$, we note that absorption cross section satisfies
 \begin{equation}
\sigma(\omega)=\frac{4\pi e^2\beta}{c} \hat{\tilde{c}}_{{\bf v}\cdot {\bf v}}(\omega),
\end{equation}
where $\hat{\tilde{c}}_{{\bf v}\cdot {\bf v}}(\omega)$ is the Fourier transform of the Kubo-transformed velocity autocorrelation function.  The RPMD model for the time correlation in Eq. (\ref{ctvv}) thus yields
\begin{equation}
\sigma^\circ(\omega)=\frac{4\pi e^2\beta}{c}\int_{-\infty}^{\infty}dt\ e^{-i\omega t} \tilde{c}_{{\bf v}\cdot {\bf v}}(t).
\label{ftctvv}
\end{equation}
 
 Finally, the parameter $\alpha$ in Eq. (\ref{Qopt}) was obtained using the L-curve method.\cite{Law74,Mil70}
   In this method, $Q(\textbf{s}; \alpha)$ is maximized for a series of $\alpha$.  For each value of $\alpha$, the quantity $\textrm{ln}[\chi^2(\textbf{s})]$ is plotted against $\textrm{ln}[-S(\textbf{s})]$, and the resulting curve exhibits a characteristic ``L'' shape.  The bend in this curve marks the transition from a regime in which deviations from the prior model are rewarded with substantial improvements in the degree to which the solution fits the imaginary time data, to a regime in which further deviations from the prior model yield smaller improvements in the fit of the imaginary time data.  Results are reported at the value of $\alpha$ that coincides with this transition (i.e., the point of maximum curvature in the L-curve).
  
  
    We emphasize that analytical continuation can only provide a quality check for the short-time RPMD dynamics, because it offers little information about timescales that exceed $\beta\hbar$.
   This well known limitation of analytical continuation can be illustrated by considering the degree to which different dynamical timescales are weighted by the kernel $K(\omega,\tau)$ in Eq. (\ref{inveq}).
For $0<\tau<\beta\hbar$ and $t>0$, the Fourier transform of the kernel is
     \begin{eqnarray}
    \hat{K}(t,\tau)&=&\int_{-\infty}^{\infty}dt\ e^{-i\omega t} K(\omega,\tau)\nonumber\\
    &=&2\sum_{j=1}^\infty \left(1-\textrm{cos}(\frac{2\pi j\tau}{\beta\hbar})\right) \textrm{exp}\left(-\frac{2\pi j t}{\beta\hbar}\right)/j.
    \end{eqnarray}
    This result, which is obtained by straightforward contour integration after observing that the $K(\omega,\tau)$ exhibits a string of simple poles for purely imaginary values of $\omega$, shows that the degree to which the kernel reports on large $t$ vanishes exponentially.  For each value of $t$, the largest contribution to $\hat{K}(t,\tau)$ occurs at $\tau=\beta\hbar/2$, where we obtain the closed form expression
    \begin{equation}
    \hat{K}(t,\beta\hbar/2) =2\ \textrm{ln}\left[\textrm{coth}\left(\frac{\pi t}{\beta\hbar}\right)\right].
    \label{meac_shorttime}
    \end{equation}
  This expression decays exponentially at long times with a rate of $\beta\hbar/(2\pi)$.  The same rate for exponential decay was reported by Habershon \textit{et al.} for analytical continuation involving a different kernel.\cite{Hab07}

\subsection{Simulation details}

  All simulations are performed using $N=1000$ classical helium atoms in a cubic simulation cell with periodic boundary conditions.  The electron is represented using $n=1024$ ring polymer beads.  The simulation temperature is $309$ K, and the helium fluid is studied at reduced densities of  $\rho^*=0.1, 0.3, 0.5, 0.7, \textrm{and}\ 0.9$.  The values of $n$ and $N$ that we employ are slightly larger, but comparable, to the values used in previous path integral Monte Carlo simulations of this system under the same conditions.\cite{Cok87,Gal94,Gal96}  The classical limit for the dynamics of the electron is obtained by setting $n=1$.
  
  RPMD trajectories are performed by integrating Eq. (\ref{mixedEOM}).  Because of the range of timescales in the problem, some care is needed to do this integration efficiently.  The multiple time-stepping (MTS) algorithm is employed to overcome the large difference in timescale for the electronic and helium degrees of freedom.\cite{Mar96,Fre02}  In all simulations, the classical helium atoms are evolved 
 with a timestep of $0.32$ fs; during each of these nuclear timesteps, the coordinates and forces of the ring polymer for the electron are updated $170$ times.  As in previous RPMD simulations, each timestep for the electron ring polymer involves coordinate updates due to (1) forces from the physical potential ($-\nabla_{{\bf q}^{(k)}} U({\bf q}^{(k)},\bar{{\bf Q}}_1,\ldots,\bar{{\bf Q}}_N)$) and (2) the exact evolution of the purely harmonic portion of the ring polymer potential.  The combined integration scheme is time-reversible and symplectic.\cite{Tuc92}
  
All interactions are truncated at a cutoff distance of $2.5\ \sigma_{\textrm{He}}$.  Because of our MTS integration scheme, the primary computational expense arises from electron-helium force evaluations.  However, these force evaluations are easily parallelized to yield a considerable ($\sim$20-fold) speedup in the wall-clock time needed for the simulations.

Prior to performing the recorded RPMD trajectories, the system is equilibrated at each density as follows.  First, the helium fluid is equilibrated in the absence of the excess electron using a long, thermostated molecular dynamics trajectory at $309$ K.  Then, the electron ring polymer is inserted into the equilibrated configuration of the helium fluid, and the combined system is run for $200$ RPMD trajectories of length $380$ fs.   Between trajectories, the position coordinates for the system are not changed, but the momenta are resampled from the Maxwell-Boltzmann distribution.  These equilibration RPMD trajectories are discarded.

The recorded RPMD trajectories are continued from the end of the equilibration trajectories.  At each density, we perform $30$ recorded trajectories of length $7.66$ ps.  Each recorded trajectory includes $n_t=24000$ MTS integration timesteps.
As before, the momentum coordinates are resampled between trajectories.
Eqs. (\ref{covar}), (\ref{ctvv}), (\ref{ctr2}), and (\ref{pimd})  are employed during these recorded trajectories to sample the quantities of interest.

The RPMD results for $R^2(\tau)$, $C_{ij}$, and $\sigma^\circ(\omega)$ provide the input to the MEAC calculation.  The imaginary time correlation function $R^2(\tau)$ is known on the discrete set of points $\{\tau_j\}$, where $\tau_j=\frac{j}{n}\beta\hbar$ and $j=1,2,\ldots,n$.  Likewise, the covariance matrix $C_{ij}$ for the imaginary time correlation function is obtained on $i,j=1,2,\ldots,n$.   Using Eq. (\ref{ftctvv}), the RPMD prior spectrum $\sigma^\circ(\omega)$ is obtained by Fourier transforming the RPMD velocity autocorrelation function $\tilde{c}_{{\bf v}\cdot {\bf v}}(t)$ on the discrete set of points $\{\omega_k\}$, where $\omega_k=\frac{k}{\mathcal{N}}\omega_{\textrm{max}}$, $\hbar\omega_{\textrm{max}}=6.25$ eV, $k=0,1,\ldots,\mathcal{N}$, and $\mathcal{N}=n_t/2$.

Unphysical noise can be eliminated from a spectrum by damping the corresponding time correlation function at large times before performing the discrete Fourier transform.  
Thus, in preparing $\sigma^\circ(\omega)$ for the MEAC calculation, we damp $\tilde{c}_{{\bf v}\cdot {\bf v}}(t)$ at times greater than $t_{\textrm{d}}$ 
 using the weighting function
\begin{equation}
w(t) = \textrm{exp}[-(t-t_{\textrm{d}})^2/(2(\beta\hbar)^2)],
\end{equation}
where $t_{\textrm{d}}=2\beta\hbar\approx 50$ fs for $\rho^*=0.1$, and $t_{\textrm{d}}=\beta\hbar\approx 25$ fs for $\rho^*=0.3-0.9$.
The MEAC calculations were also performed using larger values of $t_{\textrm{d}}$ to confirm that our conclusions are independent of this parameter.


\section{Results and Discussion}



\subsection{Dynamics on short timescales}

\begin{figure} [h!] 
\hspace*{2cm} \includegraphics[angle=180,width=15cm]{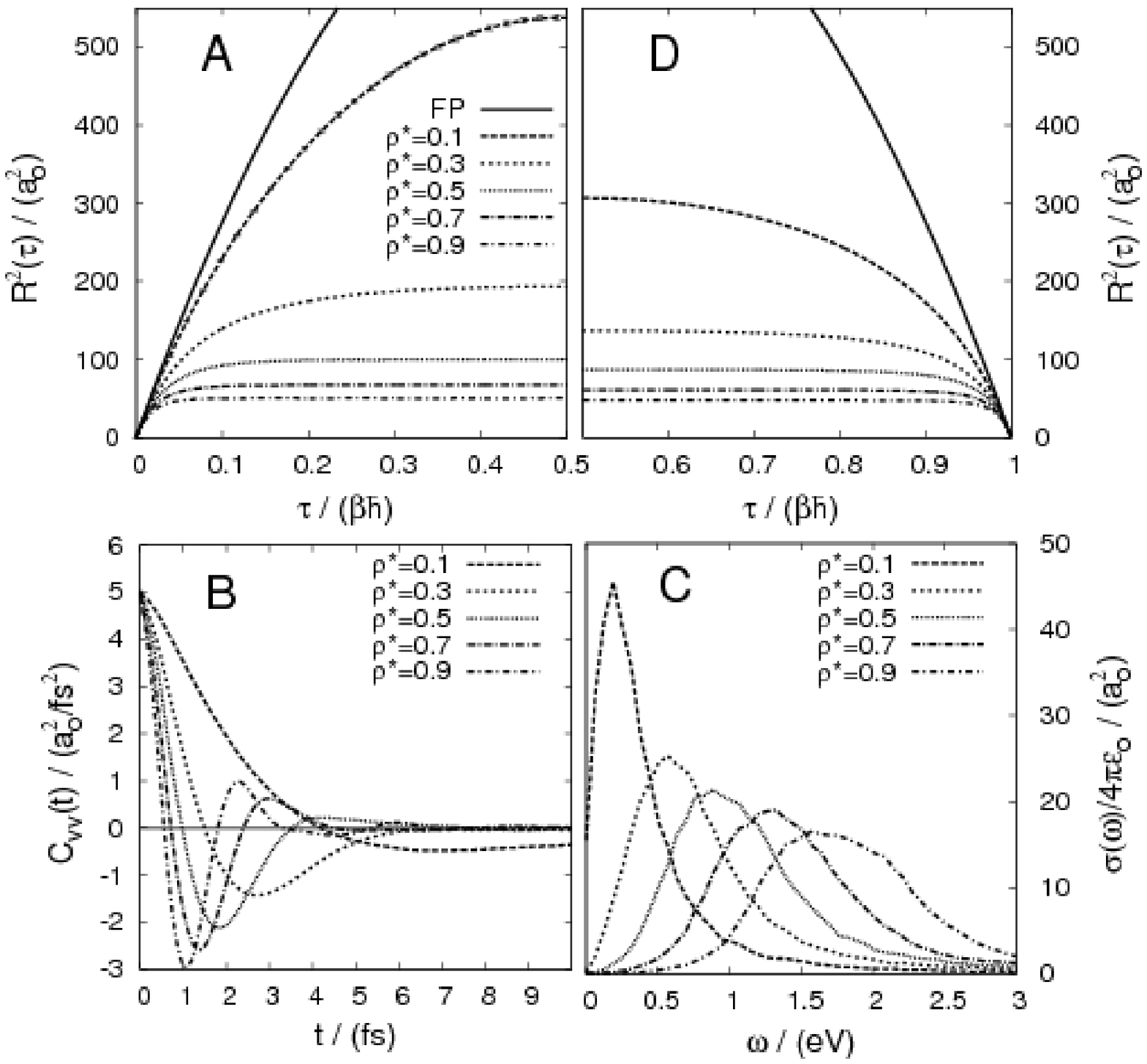}\\
  \vspace{3cm}\mbox{}
 \caption[sqr]
 {\label{fig:ctim}}
\end{figure}

Fig. \ref{fig:ctim}A shows the imaginary time correlation function $R^2(\tau)$ that is sampled during the RPMD trajectories at various densities using Eq. (\ref{pimd}).  These numerically exact results
are equivalent to previously reported path integral Monte Carlo simulations for the system considered here,\cite{Cok87,Gal94,Gal96}  
and they illustrate the finite-temperature manifestation of Anderson localization for an excess electron in a disordered medium.\cite{And58,Cha94}

For a given solvent configuration, the electron energy eigenfunctions are either spatially extended, in which they percolate across the periodic simulation cell, or they are spatially localized 
 within the cell.\cite{Cok88}
The relative stability of extended vs. localized wavefunctions hinges on a quantum mechanical balance between the penalty for delocalizing the electron wavefunction in a quasi-disordered medium and the penalty for creating a solvent cavity that is large enough to confine the localized electron.\cite{Cha94}
This balance shifts as a function of fluid density.

Since $R^2(\beta\hbar/2)$ is a measure of the spatial extent of the excess electron, the trend of increased localization with increasing fluid density is clearly seen in Fig. \ref{fig:ctim}A.
For $\rho^*=0.1$, the environment of the electron is only weakly perturbed from that of a free particle.
The fluid assumes configurations in which thermally accessible electron eigenstates are delocalized, giving rise to an imaginary time correlation function that approaches the free particle result,
\begin{equation}
R^2_{\textrm{free}}(\tau)=3\lambda^2\tau(\beta\hbar-\tau)/(\beta\hbar)^2,
\end{equation}
where $\lambda=(\hbar^2\beta/m)^{1/2}$ is the thermal wavelength of the electron.  As $\rho^*$ increases, the atoms in the fluid create a more densely disordered environment that destabilizes spatial coherence in the electron eigenfunctions.  The fraction of thermally accessible solvent configurations with thermally accessible delocalized electronic states diminishes, and the system experiences a thermodynamic driving force to localize the electron in a solvent cavity, i.e. to form a polaron.

Fig. \ref{fig:ctim}B shows the velocity autocorrelation functions $\tilde{c}_{{\bf v}\cdot {\bf v}}(t)$  that are obtained from direct RPMD simulations of the excess electron using Eq. (\ref{ctvv}).  Like the imaginary time correlation functions, these real-time correlation functions depend strongly on the solvent density.  
For $\rho^*=0.1$, $\tilde{c}_{{\bf v}\cdot {\bf v}}(t)$ decays relatively slowly and exhibits only a weak rebound due to collisions with the dilute solvent environment.  However, with increasing helium density, the ballistic motion of the electron in the RPMD model decays much more rapidly.  The timescale for the first collisional rebound of the electron (i.e., the first minima of  $\tilde{c}_{{\bf v}\cdot {\bf v}}(t)$) decreases from $7$ fs at $\rho^*=0.1$ to $1$ fs at $\rho^*=0.9$.  Furthermore, as the density approaches $\rho^*=0.9$, we see the emergence of multiple peaks in the correlation function, or ``rattling "  of the excess electron in its solvent cage.

In Fig. \ref{fig:ctim}C, we report the RPMD estimates for the absorption cross section of the solvated electron $\sigma^\circ(\omega)$, which are obtained from the $\tilde{c}_{{\bf v}\cdot {\bf v}}(t)$ using Eq. (\ref{ftctvv}).  With increasing helium density, the absorption peak is shifted to higher frequency, indicating an increasing energy gap between the ground-state and lowest-energy excited states of the electron.  Greater solvent density also broadens the spectra and elongates the high-frequency spectral tail.  All of these features are consistent with previous reports using the RISM-polaron theory for the excess electron in an adiabatic hard-sphere fluid\cite{Nic87} and using numerical analytical continuation calculations for the system considered here.\cite{Gal94,Gal96}
The only qualitative difference between the RPMD absorption cross sections in Fig. \ref{fig:ctim}C and those of previous theoretical studies\cite{Nic87,Gal96} occurs in the low density regime, at $\rho^*=0.1$.  The RPMD model predicts a maximum in the absorption cross section at finite frequency, whereas previous calculations have found that the low-density spectrum decays monotonically from a maximum at $\omega=0$.  

Eq. (\ref{inveq}) shows that the imaginary time correlation function places a constraint on the dynamics of the solvated electron.  
The accuracy of the RPMD model is thus indicated by the degree to which it violates this dynamical constraint.  In Fig. \ref{fig:ctim}D, we plot the RPMD prior estimate for the imaginary time correlation function, which is obtained by inserting $ \sigma^\circ(\omega)$ into the right-hand side of Eq. (\ref{inveq}).  
If RPMD provided an exact description of the solvated electron dynamics, then Fig. \ref{fig:ctim}D would precisely mirror the exact imaginary time correlation functions in Fig. \ref{fig:ctim}A.  
In the high density regime, this is very nearly the case; the RPMD model dynamics are consistent with the imaginary time data for $\rho^*\ge 0.5$.  However, at $\rho^*=0.1$, and to a lesser extent at $\rho^*=0.3$, the RPMD estimate deviates substantially from the exact result.  In this low-density regime, the RPMD prior spectrum $\sigma^\circ(\omega)$ corresponds to the dynamics of a model electron that is more localized than the actual solvated electron.

Since the imaginary time correlation function can be written in the basis of electronic eigenstates as follows,
\begin{equation}
R^2(\tau_k)=\frac{2}{Z}\sum_{n,m}e^{-\beta E_n}|\bra{n}{\bf q}\ket{m}|^2(1-e^{-\tau(E_m-E_n)\hbar}),
\end{equation}
we see that the changes in the value of this function with $\tau$ arise from the accessibility of excited electronic states.  At low helium densities, the fact that the imaginary time correlation function for the RPMD prior model (Fig. \ref{fig:ctim}D) plateaus at smaller values of $(\tau/\beta\hbar)(1-\tau/\beta\hbar)$ than the numerically exact result (Fig. \ref{fig:ctim}A) suggests that the RPMD model underestimates the role of these excited states in the dynamics of the electron.  By the same argument, the dominance of the electronic ground state in the dynamics at higher helium densities is correctly captured by the RPMD model.





Eq. (\ref{inveq}) can also be used to obtain a first-order correction to the RPMD prior spectrum $\sigma^\circ(\omega)$.
We employ the maximum entropy analytical continuation (MEAC) technique to numerically invert Eq. (\ref{inveq}), using $\sigma^\circ(\omega)$  as an initial guess for the solution $ \sigma(\omega)$.  As was explained in Sec. \ref{sec:meac}, this approach alters the RPMD prior spectrum to yield a solution that reproduces the imaginary time correlation functions in Fig. \ref{fig:ctim}A to within statistical certainty.



\begin{figure} [h!] 
\includegraphics[angle=180,width=18cm]{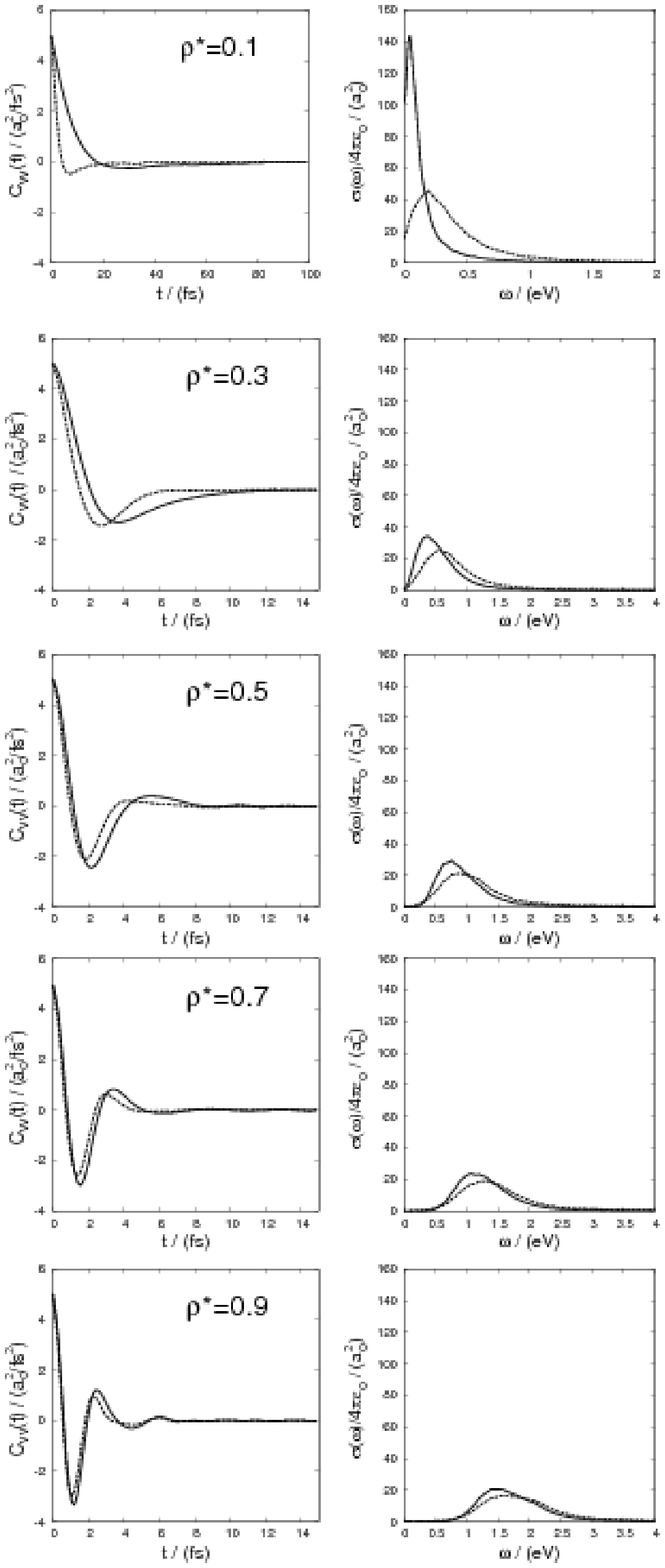}\\
  \vspace{0cm}\mbox{}
 \caption[sqr]
 {\label{fig:meac}}
\end{figure}

Fig. \ref{fig:meac} presents  the MEAC correction to the RPMD model at each density.
The dashed lines in the left column show the RPMD correlation functions $\tilde{c}_{{\bf v}\cdot {\bf v}}(t)$, and the dashed lines in the right column show the RPMD prior spectra $\sigma^\circ(\omega)$.  These results are unchanged from Figs. \ref{fig:ctim}B and \ref{fig:ctim}C, respectively. The solid lines in the right column present 
the MEAC corrections to the RPMD prior spectra. 
Finally, the solid lines in the left column show the velocity autocorrelation functions that correspond to the MEAC-corrected spectra. 


Both columns of Fig. \ref{fig:meac} illustrate that 
the MEAC procedure alters the RPMD model less with increasing helium density.  
This correction generally narrows, red-shifts, and intensifies the absorption band for the RPMD prior.
For $\rho^*=0.1$, the MEAC correction is most dramatic, causing a substantial increase in the absorption intensity and a shift  in the absorption peak to nearly $\omega=0$. 
This result again suggests that the RPMD model underestimates the role of low-lying electronic excited states in the dynamics of the electron at low solvent densities.
For $\rho^*>0.3$, the MEAC correction is much reduced, and it no longer changes the qualitative features of the RPMD prior.  Indeed, for $\rho^*=0.7$ and $0.9$, the RPMD time correlation function and the MEAC-corrected results are very similar.  


It is not surprising that the RPMD model is more reliable at high densities than at low densities.  In the  low density regime, where the electron achieves high mobility via thermal access to extended electronic states, the dynamics of the electron is a coherent scattering problem that is beyond the realm of applicability for the simple RPMD model.  On the other hand, at higher densities, the dynamics of the electron is almost entirely ground-state dominated and governed by the electronically adiabatic diffusion of the solvent-electron polaron.  In this case, since RPMD exactly describes both the structural (i.e. static) features of the localized electron and the classical dynamics of the helium solvent, it is expected that the RPMD model is accurate.  From Fig. \ref{fig:meac}, we are encouraged that the MEAC correction to the RPMD model is minor for the densities of typical liquids. 

The MEAC-corrected spectra reported in the right column of Fig. \ref{fig:meac} can be compared with previous MEAC studies for the same system.\cite{Gal94,Gal96}
The primary way in which our calculations differ from this earlier work is in the choice of prior spectrum for the maximum entropy inversion; we use the RPMD model to obtain the prior spectrum, whereas Ref. \onlinecite{Gal94} employed a flat prior spectrum and Ref. \onlinecite{Gal96} employed a prior spectrum that is based on sum rule constraints on the imaginary time correlation function.
Differences in the results of these three studies illustrate that, given the finite statistical error in the calculation of the imaginary time correlation function, the maximum entropy inversion of Eq. \ref{inveq} is unable to fully overcome different choices for the prior spectrum.  Nonetheless, we find that the MEAC technique is a useful means of testing the short timescale description of simple dynamical models such as RPMD.

\subsection{Dynamics on long timescales}

\begin{figure} [h!] 
\hspace*{0.5cm} \includegraphics[angle=180,width=18cm]{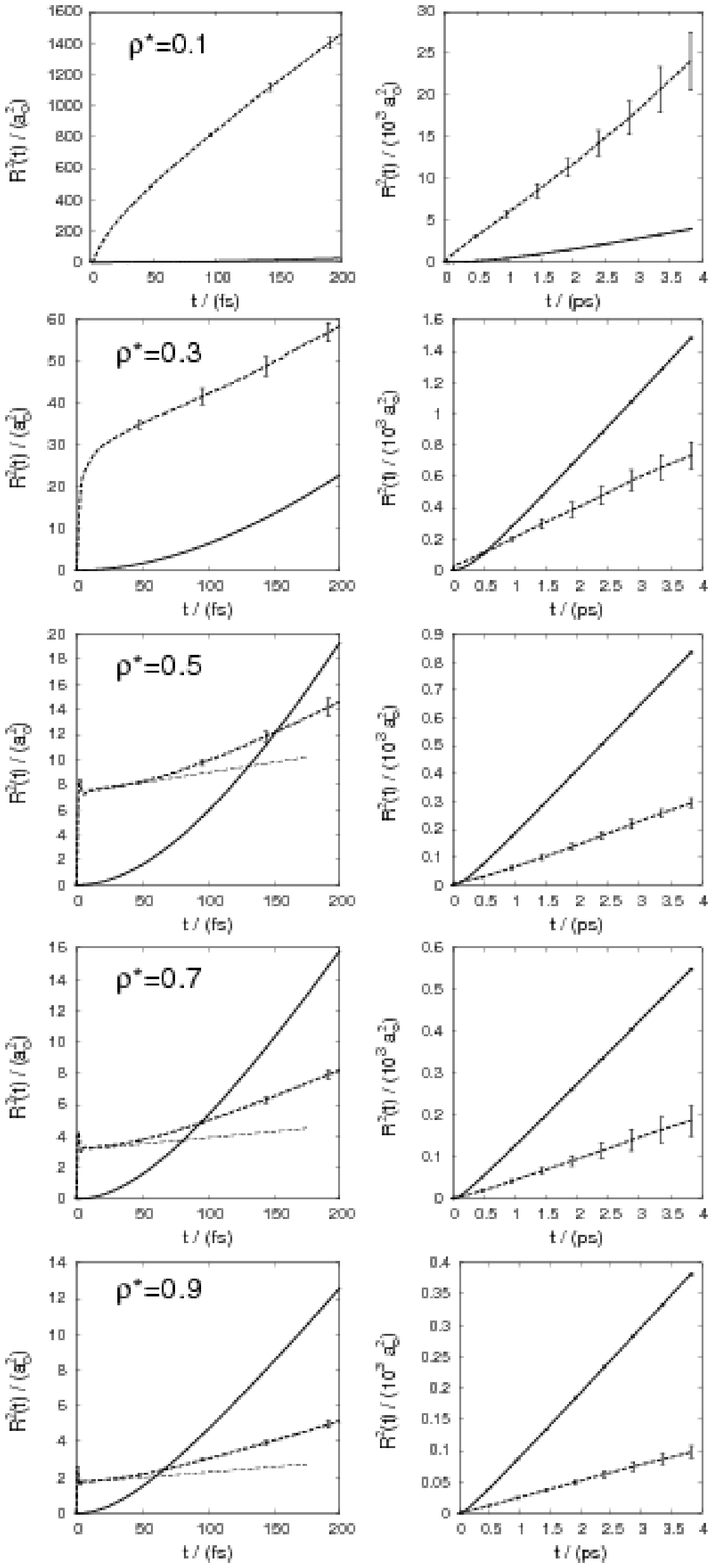}\\
  \vspace{0cm}\mbox{}
 \caption[sqr]
 {\label{fig:r2}}
\end{figure}

The left column of Fig. \ref{fig:r2} presents the mean squared displacement (MSD) for the electron using the RPMD model (dashed lines) for times up to $200$ fs.  For comparison, we also include the MSD for the atoms in the helium solvent (solid line), and at higher densities, we plot the linear fit to the RPMD curve at  $t=25$ fs, (i.e., $t=\beta\hbar$).
It is clear, especially at higher densities, that the dynamics of the electron does not reach the diffusive regime on the timescale of $\beta\hbar$.  
The MSD curves continue to deviate from linear behavior for many times this thermal timescale.
This is at first surprising, since the RPMD velocity autocorrelation functions appear to decay quickly in Fig. \ref{fig:meac},  
but the apparent contradiction simply illustrates that the self-diffusion of the electron is coupled to the slower dynamics of the solvent atoms.  
This is easily understood in the high density limit, where the delocalized states of the electron are thermally inaccessible from the strongly localized ground state and the mobility of the electron arises almost exclusively from the non-adiabatic diffusion of the solvent-electron polaron.  In this limit, the mobility of the electron would approach exactly zero if the solvent were fixed,\cite{Nic87} so any observed diffusion of the electron must be coupled to movements in the solvent environment.   
For all $\rho^*>0.5$, the left column of Fig. \ref{fig:r2} suggests that coupling of the electronic and solvent dynamics plays an important role.
The rapid decay of the velocity autocorrelation functions in Fig. \ref{fig:meac} is thus a graphical illusion; as is required by the mathematical equality in Eq. (\ref{r2vsctvv}), integration of the long-time tail reveals its non-zero contribution to the slope of the mean-squared displacement.

The importance of dynamical timescales that are slow in comparison to $\beta\hbar$ also deserves commentary from a methodological perspective.  As we have already pointed out in connection to Eq. (\ref{meac_shorttime}), the MEAC technique contains exponentially little information about dynamics on these long timescales, which explains why previous theoretical studies of the solvated electron have not observed the slow onset of diffusive behavior.\cite{Gal94,Gal96}  Furthermore, the importance of long timescale dynamics underscores the need for a model dynamics that rigorously preserves the quantum Boltzmann distribution, such as RPMD or centroid molecular dynamics.  For methods such as the linearized semiclassical initial value representation (SC-IVR),\cite{Wan98,Sun98,Liu07} in which classical molecular dynamics trajectories are initialized from a quantum distribution, it is expected (because of the negligible mass of the electron in comparison to the combined mass of the solvent) that the long-time asymptote of the electron dynamics will be independent of the initial distribution and will incorrectly converge to a purely classical description.  Conclusions about the coupled dynamics of the electron and solvent may then be obscured by the gradual breakdown of the statistical description with time.\cite{Vot05}

The right column of Fig. \ref{fig:r2} presents the RPMD results for the electron MSD on the timescale of picoseconds (i.e., up to $\approx 150\beta\hbar$).  Again, we include the solvent MSD for comparison.
The straight lines in the left column of Fig. \ref{fig:r2} indicate that both  the electron and the solvent have reached the diffusive regime.
The slope of the dashed line can be used to obtain the self diffusion coefficient for the RPMD model of the electron dynamics,
\begin{equation}
D_{\textrm{RP}}=\frac{1}{6}
\lim_{t\rightarrow\infty}\frac{d}{dt}R^{2}(t).
\end{equation}
Eq. (\ref{r2vsctvv}) ensures that this expression for $D_{\textrm{RP}}$ is equivalent to evaluating the integral of the RPMD velocity autocorrelation functions.



In Table \ref{tab:diffcoef}, we report the RPMD electron self-diffusion coeffient $D_{\textrm{RP}}$ and the helium solvent diffusion coefficients $D_{\textrm{He}}$.  Also included is the self-diffusion coefficient $D_{\textrm{cl}}$ for the electron obtain from classical molecular dynamics simulations  (i.e., 1-bead RPMD).\cite{com1}
Each is obtained from a linear least-squares fit of the corresponding MSD curve between $1$ ps and $3.7$ ps. 
Although both the classical\cite{Dun93,Yeh04} and the RPMD\cite{Mil05a,Mil05b} self-diffusion coefficients are subject to a hydrodynamic system size effect, it is not likely to alter our conclusions here.\cite{com2}
Also in Table \ref{tab:diffcoef}, we include an estimate for the radius of the excess electron at the various solvent densities.  The size of the repulsive core for the electron is obtained by evaluating the distance at which the electron-helium interaction potential in Eq. (\ref{Vehe}) first goes to zero, namely $R_0=(B-C)^{1/6}\approx1.35\sigma_{\textrm{He}}$.  The effective radius of the classical model for the electron is thus $R_{\textrm{cl}}=R_0-0.5\sigma_{\textrm{He}}\approx0.85\sigma_{\textrm{He}}$.  In the RPMD model for the electron, the quantum dispersion of the electron swells its size by an amount that can be estimated from the radius of gyration of the ring polymer, such that $R_{\textrm{RP}}=R_{\textrm{cl}}+R_{\textrm{g}}$, where
\begin{equation}
R^2_{\textrm{g}}=\left\langle\frac{1}{n}\sum_{\alpha=1}^n({\bf q}^{(\alpha)}-\bar{{\bf q}})^2
\right\rangle_{\textrm{RP}}.
\end{equation}


Comparison of $D_{\textrm{RP}}$ and $D_{\textrm{cl}}$ in Table \ref{tab:diffcoef} reveals that at every density considered, inclusion of quantum effects with the RPMD model substantially decreases the mobility of the excess electron. 
This differs from the quantum effect on self-diffusion in more weakly quantum mechanical systems, such as bulk liquid water.\cite{Mil05b}
In a weakly quantum mechanical system, quantum dispersion softens the interaction potentials without greatly affecting the structure of the system or the mechanism of diffusion; the result is an enhanced rate of self-diffusion.
The highly quantized excess electron, however, presents a much larger collisional cross section to the fluid  atoms than does the corresponding classical model, resulting in greater drag and a lower self-diffusion coefficient.

It is also interesting to compare the self-diffusion coefficients for the electron and the solvent atoms in Table \ref{tab:diffcoef}.  At the lowest fluid density, the RPMD model electron diffuses much more rapidly than the solvent atoms, but the reverse is found at all higher densities.  To understand these results, we can compare the ratio of diffusion coefficients from our simulated dynamics, $D_{\textrm{RP}}/D_{\textrm{He}}$, with the Stokes-Einstein prediction for this ratio, 
\begin{equation}
\left(\frac{D_{\textrm{RP}}}{D_{\textrm{He}}}\right)\approx \frac{R_{\textrm{He}}}{R_{\textrm{RP}}},
\label{SEratio}
\end{equation}
where $R_{\textrm{RP}}$ is the calculated size of the quantized electron in Table \ref{tab:diffcoef}, and $R_{\textrm{He}}=0.5\sigma_{\textrm{He}}$ is the van der Waals radius of the solvent atoms.  Eq. (\ref{SEratio}) simplistically assumes that the electron undergoes overdamped 
diffusion as a spherical solute of fixed radius $R_{\textrm{RP}}$.

\begin{figure} [h!] 
\hspace*{0cm} \includegraphics[angle=-90,width=13.5cm]{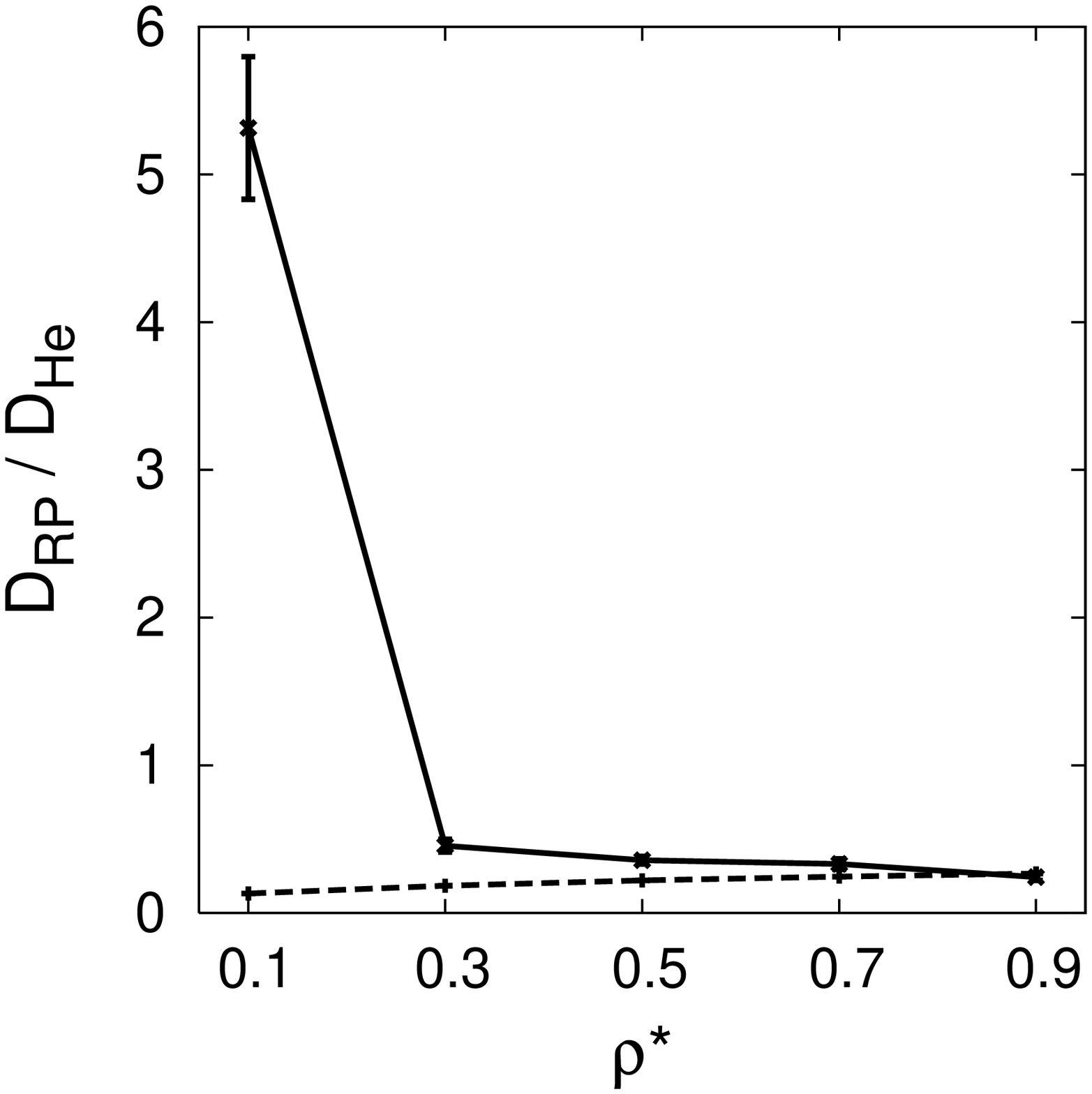}\\
  \vspace{3cm}\mbox{}
 \caption[sqr]
 {\label{fig:diffratios}}
\end{figure}

In Fig. \ref{fig:diffratios}, we plot $D_{\textrm{RP}}/D_{\textrm{He}}$ from the simulation results in Table \ref{tab:diffcoef} (solid) and from the  Stokes-Einstein model in Eq. (\ref{SEratio}) (dashed).  At high densities, the different estimates agree, 
which is reasonable given that the electron is strongly localized in this regime.  
However, for lower densities, the diffusion of the electron in the RPMD model is 
enhanced relative to the prediction of the adiabatic Stokes-Einstein model.  
This deviation from Stokes-Einstein behavior is also reasonable because of the important role of non-adiabatic transitions to delocalized electronic states in the diffusion of the excess electron at low density.
Although it was found in connection with Fig. \ref{fig:ctim} that RPMD underestimates the contribution of delocalized states to the mobility of the excess electron at low densities, we see here that the RPMD model at least partially captures this important effect.
%

\section{Conclusions}

We have used ring polymer molecular dynamics to simulate the diffusion of an excess electron in supercritical helium at various densities.  The RPMD model uses imaginary time path integrals to represent both the electron and the solvent atoms in the position basis, and it prescribes a classical molecular dynamics that rigorously samples the quantum Boltzmann distribution.  In addition to conveniently placing the electronic and solvent degrees of freedom on the same dynamical footing, the RPMD model can be used to run long trajectories and to study the coupling of different dynamical timescales.

With the aide of analytical continuation relationships, we have tested the dynamics of the RPMD model against numerically exact static correlation functions (Figs. \ref{fig:ctim}A and \ref{fig:ctim}D, Fig. \ref{fig:meac}).  These tests, which are most stringent for timescales shorter than $\beta\hbar$, suggest that RPMD is increasingly accurate as the solvent density increases.
Indeed, for $\rho^*>0.5$, the dynamics convincingly satisfies the analytical continuation constraint.  This result is encouraging for the prospect of using RPMD to study an excess electron in liquid water.  A simple estimate based on the peaks in the oxygen-oxygen radial distribution for ambient water reveals that that the liquid has an effective reduced density of $\rho^*\approx0.5-0.9$, a regime in which we find the RPMD model to be reliable.

At lower solvent densities, our analytical continuation results indicate that the RPMD model systematically underestimates the degree to which delocalized elecronic states enhance the mobility of the excess electron.  However, the failure at low densities is not absolute.  By comparing the $D_{\textrm{RP}}/D_{\textrm{He}}$ ratio from simulation with  an estimate based on the Stokes-Einstein diffusion of the excess electron, we find that the RPMD model does predict a substantial enhancement in the electron diffusion at low densities.  Even though RPMD does not fully account for the increase in electron mobility due to thermally accessible delocalized electronic states at low densities, it does qualitatively include the effect.

Finally, by running RPMD trajectories on the timescale of picoseconds, we have been able to fully bridge the timescales of the electronic and the solvent atom dynamics.  In doing so, we find that these timescales become coupled as the electron localizes with increased solvent density.  This fundamental example of complex dynamics is conveniently probed using the RPMD model, and it is clear that RPMD can be similarly employed to study the coupled dynamics of quantum mechanical and classical mechanical motions in other contexts.

\section{Acknowledgments}

The author sincerely thanks David Chandler, David Manolopoulos, and Bill Miller for helpful conversations.  This work was supported  by the Director, Office of Science, Office of Basic
Energy Sciences, Chemical Sciences, Geo-sciences, and Biosciences Division,
U.S. Department of Energy under Contract No. DE-AC02-05CH11231.
 Supercomputing resources were provided by NERSC.

\newpage
\noindent
{\bf TABLES}
\vspace*{1cm}

\begin{table}[h!]
 \caption[sqr]
{\label{tab:diffcoef} Self-diffusion coefficients$^a$ for the electron and the solvent helium atoms and the radius of the solvated electron$^b$ at various densities.}
\begin{tabular}{c c c c c}
\hline
\hline
$\quad\quad \rho^* \quad\quad $& $\quad\quad D_{\textrm{RP}}\quad\quad$& $\quad\quad D_{\textrm{cl}}\quad\quad$ &$\quad\quad D_{\textrm{He}}\quad\quad$ &$\quad\quad  R_{\textrm{RP}}\quad\quad$\\
\hline
0.1 & 1.1(1)        & 6.6(3)        & 0.207(4)    & 3.69 \\
0.3 & 0.032(3)   &  0.53(3)     & 0.0702(6)  & 2.71 \\
0.5 & 0.014(1)   &  0.056(3)   & 0.0391(4)  & 2.26 \\
0.7 & 0.0085(9) &  0.022(1)   & 0.0256(4)  & 2.03 \\
0.9 & 0.0043(4) &  0.0119(6) & 0.0176(2)  & 1.87 \\
\hline
\hline
\end{tabular}
\end{table}
\noindent
$^a$All self-diffusion coefficients are reported in units of $a_0^2$/fs.  The uncertainty in the final digit is indicated in parenthesis.\\
$^b$The radius of the quantized electron, $R_{\textrm{RP}}$, is reported in $\sigma_{\textrm{He}}$.  The radius of the electron in the classical model is $R_{\textrm{cl}}=0.85 \sigma_{\textrm{He}}$, and the radius of the solvent helium atoms is $0.5 \sigma_{\textrm{He}}$.  See text for details.

\newpage
\noindent
{\bf FIGURE CAPTIONS}
\vspace*{1cm}\mbox{}\\
\textbf{Figure 1.}  Data from the RPMD simulations of the excess electron in helium at various densities.
(\textbf{A}) Numerically exact imaginary time correlation functions $R^2(\tau)$. 
(\textbf{B}) RPMD velocity autocorrelation functions $\tilde{c}_{{\bf v}\cdot {\bf v}}(t)$. 
(\textbf{C}) RPMD prior spectra $\sigma^\circ(\omega)$. 
(\textbf{D}) RPMD prior estimates for the imaginary time correlation functions, which are obtained by inserting the $\sigma^\circ(\omega)$ into Eq. (\ref{inveq}).  The imaginary time correlation functions in parts A and D are symmetric about $\tau=\beta\hbar/2$.
\mbox{}\\
\textbf{Figure 2.} The MEAC correction to the RPMD model at various densities.  The left column presents the RPMD (dashed) and MEAC-corrected (solid) velocity autocorrelation functions for the electron.  The right column presents the corresponding absorption cross sections.
\mbox{}\\
\textbf{Figure 3.} The RPMD mean squared displacments for the electron (dashed) and the solvent helium atoms (solid) at various densities.  Results are presented for both short timescales (left column) and long timescales (right column).
\mbox{}\\
\textbf{Figure 4.} Ratio of the electron and helium atom self-diffusion coefficients at various densities, obtained from the RPMD simulations (solid) and from the Stokes-Einstein model in Eq. (\ref{SEratio}) (dashed).  

\newpage
\noindent
{\bf FIGURES}
\vspace*{1cm}


\newpage

\newpage

\newpage



\begin{thebibliography}{99}
\bibitem{com1} Estimates of the $D_{He}$ obtained during the RPMD simulations of the electron (see Tab. \ref{tab:diffcoef}) and during the purely classical simulations of the solvated electron (not shown) agree to within $2\%$ at each density.  The small difference in these results is a finite system size effect; since the RPMD model correctly includes quantum dispersion of the electron, it is effectively larger than in the classical simulations.  This small difference in the size of the electron changes the available space available to the solvent atoms, and thus slightly effects the solvent self-diffusion coefficient.
\bibitem{com2} We find another minor system size effect at the lowest fluid density of $\rho^*=0.1$.  Given the very high mobility of the electron at this density, the velocity autocorrelation function of the helium atoms does not fully decay to zero on the timescale required for the electron to travel the sidelength of the periodic box.  The electron can thus encounter helium atoms with which it is dynamically correlated.
\end{thebibliography}
\end{document}